\documentclass[preprint,trackchanges]{aastex62}
\usepackage{graphicx}
\hypersetup{linkcolor=blue,citecolor=blue,filecolor=cyan,urlcolor=blue}
\shorttitle{Filament Eruption and Its Reformation }
\shortauthors{Yang et al.}

\begin{document}

\title{Filament Eruption and Its Reformation Caused by Emerging Magnetic Flux}

\correspondingauthor{Bo Yang}
\email{boyang@ynao.ac.cn}

\author{Bo Yang}
\affil{ Yunnan Observatories, Chinese Academy of Sciences, 396 Yangfangwang, Guandu District, Kunming, 650216, People's Republic of China}
\affiliation{Center for Astronomical Mega-Science, Chinese Academy of Sciences, 20A Datun Road, Chaoyang District, Beijing 100012, People's Republic of China}
\affiliation {Key Laboratory of Solar Activity, National Astronomical Observatories of Chinese Academy of Science, Beijing 100012, People's Republic of China }


\author{Huadong Chen}
\affiliation {Key Laboratory of Solar Activity, National Astronomical Observatories of Chinese Academy of Science, Beijing 100012, People's Republic of China }
\affiliation{School of Astronomy and Space Science, University of Chinese Academy of Sciences, Beijing 100049, People's Republic of China}





\begin{abstract}
We present observations of the eruption and then reformation of a filament caused by its nearby emerging magnetic flux.
Driven by the emerging magnetic flux, the emerged positive fluxes moved toward and cancelled with its nearby negative fluxes,
where the negative ends of a filament channel beneath the filament and a bundle of left-skewed coronal loops overlying the filament were anchored.
Complemented by the nonlinear force-free field extrapolation, we find that the coronal magnetic field lines associated with the filament channel
and the emerging magnetic fields consist of sheared field lines. Prior to the filament eruption,
unambiguous observational evidence indicates that multiple interactions occurred between the emerging magnetic fields
and the left-skewed coronal loops, implying a tether-weakening reconnection. Specifically, during the final episode of the tether-weakening reconnection,
a remarkable sigmoid structure was formed and lifted up together with the filament. Accordingly, we speculate that
the tether-weakening reconnection probably destabilized the filament system and triggered its rise.
Subsequently, the filament and the sigmoid structure erupted together and produced a CME. After the eruption,
the emerging magnetic fields continued to reconnect with the remaining filament channel, leading to the reformation of the filament.
This observation strongly supports the idea that emerging magnetic flux plays an important role in triggering the filament to erupt,
and the filament is reformed by magnetic reconnection between the emerging magnetic fields and its nearby filament channel.
\end{abstract}

\keywords{Sun: activity -- Sun: filaments, prominences -- Sun: flares -- Sun: coronal mass ejections (CMEs) -- Sun: magnetic fields}

\section{Introduction} \label{sec:intro}
Solar filaments, which consist of cool dense plasma that protrudes high into the tenuous hot corona, are
elongated absorption features when observed on the solar disk in $H_{\alpha}$ and some extreme ultraviolet (EUV)
lines \citep{an05}. They are also known as prominences when observed in emission outside of the disk.
It is a general consensus that filament channels provide the magnetic environment in the low corona to support
the filaments against gravity and thermally isolate from the surrounding hot corona.
Filament channels are commonly thought to be the birth ground of filaments.
Previous $H_{\alpha}$ observations have demonstrated that filament channels are associated with fibril structures
aligned along the polarity inversion line (PIL) and they are locations of strong magnetic shear and
highly non-potential magnetic fields \citep{mar98,mac10}. In particular, theoretical model \citep{low95}
and observation \citep{su10} have suggested that the magnetic structure of the filament channel should be a magnetic flux rope,
which is considered to be the core structure of solar eruptions \citep{for2000}. Once the filaments erupt,
they may lead to the energetic flares and coronal mass ejections (CMEs),
which represent the main driver of space weather \citep[for reviews see, e.g.,][]{sch15,gre18}.
Understanding the mechanisms of formation, evolution, and eruption of filaments are key to our understanding of
the evolution of magnetic fields on the Sun and the initiation of solar eruptive phenomena.

The magnetic field is at the epicenter of  filaments. To reveal the detailed magnetic structures of filaments,
a lot of effort have been made to construct the models for filament magnetic fields. In general, two popular models, including
sheared arcade model \citep[e.g.,][]{ks57,ant94,dev2000,aul02,wel05} and magnetic flux rope model \citep[e.g.,][]{kr74,rus94,gib04,ama14},
have been proposed and employed to account for the magnetic structures of filaments. In both models,
the cool dense plasma of the filaments is preferentially supported in the magnetic dips of the sheared arcade
or magnetic flux rope. Many observations, which supplemented by nonlinear force-free magnetic field (NLFFF) extrapolations,
actually have confirmed such a picture \citep[e.g.,][]{can10,guo10,ama14,jiang14,bi15,yang16b}. However, in contrast to this,
other works have demonstrated that a filament is a dynamic entity and it
can be supported by sheared arcades without magnetic dips \citep{kar01,zou16}.
\citet{guo10} found in one filament that the magnetic structures of the filament was the combination of a magnetic flux rope with
dipped sheared arcades. In addition, \citet{kuc12} reported that the portion of the filament in the chromosphere was strongly sheared,
whereas the photospheric field lines underneath had an inverse polarity configuration.
Despite the detailed magnetic structures of the filaments are more complex than we thought,
these theoretical and observational studies have deepened our understanding on the magnetic structures of filaments.

How the magnetic structures of filaments are formed on the Sun is still an unanswered question in solar physics.
Based on numerous observational and theoretical works, \citet{mac10} have concluded in a review
that there are two possible mechanisms that aim to describe the formation of the magnetic structures of filaments:
the surface mechanisms \citep[e.g.,][]{van89,dev2000,mar01,wel05,ama11}
and the subsurface mechanisms \citep[e.g.,][]{rus94,low95,gib04,oka08,oka09,lit10}.
In the surface mechanisms, magnetic reconnection and photospheric motions,
such as vortical motions, shearing motions, and converging motions, play an important role in
building up the magnetic structures of filaments. The vortical motions at the magnetic footpoints of the flux tubes
can direct twist the flux tubes or bring flux systems together, leading to an eventual reconnection and
resulting in the formation of filaments \citep{yan15,yang15a,chen18b}. The shearing motions and the converging motions,
in particular shearing motions along a PIL and converging motions onto a PIL, can drive opposite-polarity magnetic
elements together and then facilitate magnetic reconnection and associated magnetic flux cancellation at the PIL,
which gradually transforms short sheared arcades into longer helical flux ropes that is capable of supporting filament plasma \citep{van89,chae01,gai97,wang07}.
Recently, an increasing number of observations have further validated the surface mechanisms and revealed more details of
the formation process of the filaments \citep{yan2016,yang16a,yar16,zhou16,kumar17,wang2017}. In particular, \citet{yang16a} found that
magnetic reconnection associated with flux convergence and cancellation can result in the rapid formation of a filament within about 20 minutes.
By conducting a three-dimensional magnetohydrodynamics (MHD) simulations, \citet{kan17} reproduced this formation process and
found that magnetic reconnection can lead not only to flux rope formation, but also to filament plasma formation via radiative condensation.

In the subsurface mechanisms, there should be a twisted flux rope generated in the convection zone and then partly emerges into
an overlying arcade by buoyancy. Simultaneously, the cool dense plasma resided in the dips of the flux rope or
U-loop is lifted by the rising twisted field lines\citep{rus94,lit97}.
Hitherto, only a handful of observations are in favor of the subsurface mechanisms. By analysing the evolution of photospheric vector magnetograms
under an AR filament, \citet{oka08,oka09} presented, for the first time, two solid observational evidences supporting
an emerging horizontal flux rope scenario. The first is that the opposite polarities of a long arcade structure move apart and then come together.
The second is that the orientations of horizontal magnetic fields along the PIL on the photosphere change from
normal to inverse polarity. In the course of the evolution of the photospheric vector fields,
\citet{oka09} found that the filament above alters its appearance from a single to a fragmented one and back again.
In addition, frequent transient brightenings are also occurred along the filament segments. Accordingly, \citet{oka09} suggested that
the emerging helical flux rope reconnects with the magnetic fields of the pre-existing filament that constructs the longer coherent filament.
This scenario has also been reported by \citet{bu16}. The observations of \citet{oka09} and \citet{bu16} implied that magnetic reconnection also
plays a key role in the subsurface mechanisms. More recently, \citet{yan2017} presented a complete process that a small-scale flux rope
emerged and then erupted, which produced an M-class flare and a CME. Motivated by \citet{oka08,oka09}, \citet{mact10}
simulated the rise of a flux rope from the solar interior into an overlying arcade and confirmed the results of \citet{oka08,oka09}.
In contrast, \citet{var12} demonstrated that the observations of \citet{oka08,oka09} can also be well explained by the flux cancellation process.
On the other hand, many numerical experiments have demonstrated that the flux rope can not be bodily lifted into the corona by only
using buoyancy and magnetic buoyancy instabilities \citep{fan01,manc04,hood12}. However, as impacted by shearing motions and rotational motions, the reconnection of
emerged sheared field lines that lie above the emerging tube axis can also lead to the formation of a flux rope \citep{arch08,fan09,arch14}.
Therefore, the subsurface mechanisms are still controversial and more detailed observations of the newly emerging flux and associated filament formation process
need to be further investigated.

Newly emerging magnetic flux plays not only an important role in the formation of filaments but also in the trigger mechanisms
for filament eruption \citep{cs00,lin001,chen18c,dac18}. \citet{fey95} and \citet{wang99} suggested that when newly emerging magnetic
flux was oriented favorably for reconnection with the overlying arcade field containing the filament, the following reconnection between
them can divert the flux overlying the filament sideways or to greater heights, triggering  the slow rise of the filament.
However, \citet{lin001} analytically investigated the interaction between the emerging magnetic flux and the pre-existing coronal magnetic field
which contains a flux rope and pointed out that no simple, universal relation between the orientation of the emerging magnetic flux and the likelihood of an eruption.
\citet{cs00} simulated an emerging magnetic flux trigger mechanism
and indicated that when reconnection-favored emerging magnetic flux emerges
within the filament channel, it cancels the magnetic fields below the flux rope, and when reconnection-favored emerging magnetic flux appears
on the outer edge of the filament channel, it rearranges the global structure of the overlying coronal field. Both different scenarios
can reduce the magnetic tension in the flux rope or overlying sheared arcades and lead to the slow rise of the filament.
The newly emerging magnetic flux interacted with the filament channel and the overlying sheared arcades
are similar to the tether-cutting \citep{moo80,moo01,ji08,liu10,chen14,chen16,xue17}
and tether-weakening schemes \citep{nag07,ster07,ster14}, respectively. In both schemes, canceling magnetic flux
usually happened on or near the PIL and soft X-ray (SXR) or EUV microflarings occurred repeatedly at the flux-emergence sites \citep{chi07,ster07,ster14}.
\citet{chi07} suggested that the microflarings were precursors to the filament eruption and they were evidence for
a tether-cutting mechanism. In the tether-weakening scheme, a solid observational evidence is the formation of longer
and higher arching loops in the overlying arcades, which display bright fanlike SXR structure stemming from the microflares \citep{moo92,ster07}.

In this paper, with high-quality data acquired by the New Vacuum Solar Telescope \citep[NVST;][]{liu01,liu14} and the \emph{Solar Dynamics Observatory}
\citep[\emph{SDO};][]{pes12}, we present the eruption and then the reformation of a filament caused by an
newly emerging magnetic flux, which emerged close to the negative ends of the filament. This enables us to investigate the trigger and formation mechanisms
of the filament in detail. The remainder of the paper is organized as follows: the detailed observations and methods are described in Section 2,
the main results are shown in Section 3, and a conclusion and a discussion are given in Section 4.

\section{Observations and Methods}
The eruption and then reformation of the filament occurred in NOAA active region (AR) 11791 on 2013 July 15
and were well-observed by the Atmospheric Imaging Assembly \citep[AIA;][]{lem12}
and the Helioseismic and Magnetic Imager \citep[HMI;][]{sch12} on board the \emph{SDO}. The AIA instrument takes the full-disk images of
the Sun in 10 ultraviolet (UV) and EUV wavelengths and up to 0.5 $R_\sun$ above the solar limb. The spatial resolution of the images is
1.$\arcsec$5 (0.$\arcsec$6 pixel$^{-1}$) and the time cadence of the images is up to 12 s. In this study, the 304 \AA\ (\ion{He}{2}, 0.05 MK)
, 94 \AA\ (\ion{Fe}{18}, 7 MK), and 193 \AA\ (Fe {\sc xii}, 1.3 MK and Fe {\sc xxiv}, 20 MK) images were used.
The HMI instrument measures the full-disk continuum intensity images and the line-of-sight
(LOS) magnetic field for the \ion{Fe}{1} absorption line at 6173 \AA\,.
In addition, the continuous photospheric vector field \citep{tur10} in the so-called HMI Active Region Patches (HARPs) region was also provided by the HMI.
The time cadence of the HMI are 45 s (for the continuum intensity images and the LOS photospheric magnetograms) and 720 s (for vector magnetic field data),
and the spatial resolution of the HMI is 1.$\arcsec$0. All images taken from the \emph{SDO} were then aligned by differentially rotating to the reference time of
03:10 UT on 2013 July 15.

On 2013 July 15, AR 11791 was also observed by the NVST in $H_{\alpha}$ 6562.8 \AA\ and TiO during two periods (01:15-02:18 UT; 06:10-08:19 UT).
The pixel size of the $H_{\alpha}$ and TiO images are 0.$\arcsec$163 and 0.$\arcsec$035, respectively. The time cadence of these images is 12 s.
By subtracting dark currents and correcting with flat field, the raw data were first calibrated to Level 1, and then the calibrated images were further
reconstructed to Level 1+ by speckle masking method \citep{wei77}. The high-resolution reconstruction process of the NVST data were described in detail by \citet{xiang16}.
Based on a high accuracy solar image registration procedure \citep{feng12,yang15b}, all of the NVST images were co-aligned with each other.
In the present study, the \emph{Reuven Ramaty High Energy Solar Spectroscopic Imager} \citep[\emph{RHESSI};][]{lin02}
were also utilized to investigate the hard X-ray sources related to the particle acceleration sites. We used the CLEAN method \citep{hur02}
to reconstruct the \emph{RHESSI} image at the energy bands of 6-12 and 12-25 keV, with an integration time of 2 minutes (01:54-01:56 UT; 03:01-03:03 UT; 06:39-06:41 UT).
To present the variation of soft X-ray (SXR) 1\sbond8 \AA\ flux in the course of recurrent brightenings and the filament eruption,
the observations from the \emph{Geostationary Operational Environmental Satellite} (\emph{GOES}) were also ultilized.
Moreover, observations from the Large Angle and Spectrometric Coronagraph Experiment (LASCO) \citep{bru15} on-board the \emph {Solar Heliospheric Observatory} (\emph {SOHO})
were also employed to identify the associated CMEs.

In order to obtain the magnetic field topology of the AR, the "weighted optimization" method \citep{whe2000,wie04} is used to perform a NLFFF extrapolation.
Before the extrapolation, the very fast inversion of the Stokes vector algorithm \citep{bor11}
is utilized to compute the vector field data, and the minimum energy method \citep{met2006,lek2009}
is used to resolve the remaining 180\arcdeg\ azimuth ambiguity.
To remove the projection effect, the HARP vector field data are remapped to a lambert cylindrical equal-area (CEA)
projection and then transformed into standard heliographic spherical coordinates. In addition, the bottom boundary vector data are reprocessed by
a procedure developed by \citet{wie06}. This process drives the observed non-force-free data toward suitable boundary conditions for
a force-free extrapolation.

\section{Results}
\subsection{Overview of AR 11791 and the Filament}
The AR 11791 was located at about S$15\degr$E$03\degr$ on 2013 July 15. Figure 1({\it a}\sbond{\it b}) illustrate the general appearances of the AR.
The leading polarity of the AR was composed of a few positive penumbra spots and its following polarity was mainly consisted of two negative sunspots,
labeled as ``N" and ``n1", respectively. The newly emerging magnetic flux emerged at the northeast of N and its positive polarity,
labeled as ``p", moved toward n1. As a result, magnetic flux cancellation between the opposite-polarity magnetic flux patches p and n1 will
occur when they collided with each other. While its negative polarity partly merged with N and partly moved to the east of N,
forming several negative penumbra spots labeled as ``n". The filament of interest showed up as a J-shape (panel ({\it c})),
with its positive ends apparently rooted in the leading penumbra spots ``P" and its negative ends rooted in the trailing sunspot N.
When viewed from the positive-polarity side of the filament, the axial field component of the filament points to the right.
In addition, some left-skewed coronal loops overlying the filament and connecting the leading positive penumbra spots ``p1" to n1 were also observed
in the AIA 193 \AA\ image (panel ({\it d})). According to the definition of the filament chirality \citep{mar98},
these observations demonstrate that the filament is dextral, which is in conflict with the empirical hemispheric chirality rule of filaments.

\subsection{Recurrent EUV Brightenings Prior to the Filament Eruption}
Prior to the filament eruption, the light curves of \emph{GOES} SXR flux and \emph{RHESSI} HXR flux display a total of nine impulsive peaks
from 01:20 UT to 02:20 UT (as indicated by the arrows in Figure 2({\it a})).
It is noted that each impulsive peak except the eighth is consistent with a \emph{RHESSI} HXR source at almost the same location.
Scrutinizing the EUV observations from the \emph{SDO}/AIA movie (see the animation of Figure 2),
we found that a series of EUV brightenings occurred repeatedly near the negative
ends of the filament from about 00:50 UT to 02:30 UT. The snapshots of the recurrent brightenings
and the evolution of the filament are shown in a sequence of AIA 304 \AA\ (Figure 2({\it b}\sbond{\it d})), NVST $H_{\alpha}$
(Figure 2({\it e}\sbond{\it g})), and AIA 94 \AA\ (Figure 2({\it h}\sbond{\it j})) images. When contours of HMI magnetograms were superimposed
on simultaneous AIA 304 \AA\ (Figure 2({\it c}) and 94 \AA\ (Figure 2({\it h}) images, it became clear that the recurrent EUV brightenings
were exactly originated from the region where the opposite-polarity magnetic flux patches p and n1 collided.
We also overlayed the \emph{RHESSI} HXR sources on the AIA 304 \AA\ (Figure 2({\it c})
and 94 \AA\ (Figure 2({\it i}) images and found that the \emph{RHESSI} HXR sources in the energy ranges of 6\sbond12 keV appeared
almost coincident with the strongest EUV brightenings. Accordingly, we speculate that the impulsive peaks except the eighth should be closely related
to the recurrent EUV brightenings. All of these observational features suggest that magnetic reconnection may occur between
the emerging magnetic flux and the magnetic structures that rooted in n1.

By tracking the evolution of the EUV brightenings in detail, we found two stages of interaction between
the emerging magnetic flux and the magnetic structures that rooted in n1.
The first stage of interaction occurred from about 00:50 UT to about 01:20 UT. In this time interval, we observed that the recurrent EUV brightenings
in the AIA 94 \AA\ observations firstly appeared at the opposite-polarity magnetic flux region
and then noticeably elongated along the lower edge of the filament in opposite directions.
As a result, the lower edge of the filament was progressively brightened (Figure 2({\it h})).
In addition, the AIA 304 \AA\ observations (Figure 2({\it b})) show that
hot material also  originated from this region and moved along the lower edge of the filament in opposite directions.
Along slice ``AB" marked in Figure 2({\it b}), a space-time plot was constructed from AIA 304 \AA\ images
and the result was provided in Figure 4({\it a}). The hot material spread along the lower edge of the filament
in opposite directions is clearly shown on the space-time plot.
Via performing linear fittings to the outer edge of the hot material, it is found that the hot material moved to the positive ends of the filament
at a projected velocity of about 80.0 km s$^{-1}$ and to the negative ends of the filament at about 74.5 km s$^{-1}$.
In the chromosphere, the NVST $H_{\alpha}$ image (Figure 2({\it e})) shows that footpoint brightenings appeared at the positive footpoints of the
filament P, the negative footpoints of the left-skewed coronal loops n1, and the footpoints of the emerging magnetic flux (p and n).
Based on these observations, we speculate that magnetic reconnection took place in this stage is very likely between the emerging magnetic flux and
the magnetic structures, which should be the fibril structures of the filament channel connecting P to n1 and underneath the filament.
The magnetic reconnection occurred in this stage shows some similarities with the tether-cutting scheme \citep{moo01,chi07,ji08,chen16,xue17}.

The second stage of interaction occurred after about 01:20 UT and continued until the eruption of the filament.
Likewise, recurrent EUV brightenings and hot material that originated in the
opposite-polarity magnetic flux region appeared again. However, the hot material does not spread along the filament in opposite directions
but along a trajectory consistent with the position of the left-skewed coronal loops (Figure 2({\it c})).
In the AIA 94 \AA\ observations, the left-skewed coronal loops were progressively illuminated and bright fanlike structures stemming from the
source region of the EUV brightenings were formed (Figure 2({\it i}\sbond{\it j})).
In the chromosphere, the NVST $H_{\alpha}$ images (Figure 2({\it f}\sbond{\it g})) show
that footpoint brightenings mainly appeared at the footpoints of the the emerging magnetic flux (p and n) and
the left-skewed coronal loops (p1 and n1). Apart from that, dark fibrils striding the filament and rooting in p1 and N were observed
in the $H_{\alpha}$ image (Figure 2({\it f})). These observations  provide solid evidence that magnetic reconnection happened
between the emerging magnetic flux and the left-skewed coronal loops overlying the filament. The reconnection between them produced the
recurrent EUV brightenings, the material flows, the footpoint brightenings, and heated the plasma on the reconnected field lines
forming the bright fanlike structures. Moreover, the reconnection may divert the negative footpoint of the left-skewed coronal loops
to N, producing the dark fibrils striding the filament and rooting in p1 and N. The magnetic reconnection occurred in this stage is
indicative of the tether-weakening scheme \citep{moo92,nag07,ster07}, which will
lead to the slow rise of the filament.

\subsection{Eruption of the Filament}
Figure 3 shows the eruption process of the filament. Shortly before the rise of the filament at about 02:45 UT, a more intense EUV brightening appeared
at the opposite-polarity magnetic flux region (panels ({\it a}) and ({\it e})). This intense EUV brightening is characterised by a C1.0 flare, which starts at 02:33 UT,
peaks at 02:57 UT, and ends at 03:03 UT registered by \emph{GOES}. During the flare, a remarkable sigmoid structure (panels ({\it f})\sbond ({\it g})),
which manifests itself as a bright inverse S-shaped hot channel \citep{zhang12,cheng13}, was gradually formed.
However, in this study, the inverse S-shaped hot channel is not strictly along the PIL.
It connects p1 to n and strides over the filament (as indicated by the composite image in panel ({\it j})). Similarly, a \emph{RHESSI} HXR source
in the energy ranges of 6\sbond12 keV was also found spatially coincident with the EUV brightenings (panel ({\it f})).
These features indicate that tether-weakening reconnection further occurred between the emerging magnetic flux and the left-skewed coronal loops,
which results in the flare and the formation of the inverse S-shaped hot channel. In the course of the flare,
the filament was observed to rise slowly (panels ({\it b})\sbond ({\it d})). Meanwhile, an inspection of the movie (see the animation of Figure 3) reveals
that filament materials moved to both the positive and negative footpoints of the filament. Accompanying with the slow rise of the filament,
the inverse S-shaped hot channel progressively expanded (panels ({\it g})\sbond ({\it h})). Subsequently, after about 03:15 UT, the filament
together with the inverse S-shaped hot channel erupted violently to the southeast and ejected out of the FOV of panels ({\it a})\sbond ({\it h}).
Panel ({\it i}) demonstrates the erupted filament in a larger FOV. It is worth noting that two dimming regions, which correspond to the footpoints of the
inverse S-shaped hot channel p1 and n, were gradually formed and identified through the AIA 193 \AA\ image (panel ({\it k})).
As suggested by many previous researches \citep{ster97,jiang03,tian12} that these dimming regions should be due to the reduction in plasma density
in the footpoints of the expanding and erupting magnetic volume, such as filament and sigmoid structure. It is registered by \emph{GOES} that
the eruption of the filament is accompanied by a C3.0 flare that starts at 03:11 UT, peaks at 03:44 UT, and ends at 04:08 UT. In addition,
as shown in the LASCO/C2 white-light base-difference image that the eruption of the filament finally produced a CME (detailed information of this CME,
see the LASCO CME catalog: \url{https://cdaw.gsfc.nasa.gov/CME_list/UNIVERSAL/2013_07/univ2013_07.html}).

To comprehensively investigate the entire eruption process of the filament, a space-time plot, which is approximately along the erupting direction
(as indicated by the long white arrow ``CD" in panel ({\it i})) of the filament, was constructed from AIA 304 images and the result was presented in Figure 4({\it b}).
One can see that the filament primarily experiences two distinct motion stages: a slow rise phase immediately followed by a sudden eruption phase.
An application of linear fittings to the outer edges of the erupting filament gives average speeds of 30.8 km s$^{-1}$ and 180.8 km s$^{-1}$
for the slow rise phase and the sudden eruption phase, respectively. It is noted from the \emph{GOES} SXR flux that a small hump, which corresponds
to the peak of the C1.0 flare just before the onset of the main C3.0 flare, appeared. The rise of the filament occurred after the onset of the
C1.0 flare and continued till the initiation of the C3.0 flare, and the sudden eruption of the filament occurred during the C3.0 flare.
\citet{shen12} have reported a similar observation and they suggested that their observation supports a magnetic breakout scenario.
Here, these observational signatures indicate that tether-weakening reconnection occurred between the emerging magnetic flux and the left-skewed coronal loops and resulted in
the C1.0 flare, which is found to be closely related in time to the slow rise phase of filament. It seems that the C1.0 flare, which produced by the final episode of
tether-weakening reconnection, was a direct trigger of the slow rise of the filament. Even though the final episode of tether-weakening reconnection
was found to be the most intense and was closely related in time to the slow rise phase of filament. However, we can not discern whether the slow rise of the filament
was mainly caused by the final episode of tether-weakening reconnection. In our observations, the tether-weakening reconnection
happened before the C1.0 flare diverted the negative footpoints of the left-skewed coronal loops to N,
while the final episode of tether-weakening reconnection diverted the negative footpoints of the left-skewed coronal loops even to n.
All of these reconnection processes play a role in lengthening the left-skewed coronal loops overlying the filaments,
which can reduce the magnetic tension force that restraining the filament, thus destroy the magnetic balance of the filament system and
lead to the rise of the filament. Therefore, we suggest that the slow rise of the filament may be triggered by a series of
tether-weakening reconnection processes.

\subsection{Reformation of the Filament}
Previous observations have reported that successive filaments may reform during or within
a few hours after the filament eruption \citep{joshi14,yan15,chen18a,cheng18}.
However, the detailed reformation process has rarely been captured. In our observation, after the filament erupted, another filament reformed at the position of the
erupted one. Figure 5 shows the reformation process of the filament in the NVST $H_{\alpha}$ images (panels ({\it a})\sbond ({\it h})) and the AIA 304 \AA\ images
(panels ({\it i})\sbond ({\it p})). After the filament erupted, a filament channel, which consists of fine chromospheric fibrils connecting P to n1,
were effectively identified by the NVST high resolution $H_{\alpha}$ image (panel ({\it a})). These chromospheric fibrils lie below
the erupted filament providing evidence to support our previous result that magnetic reconnection occurred between the emerging magnetic flux
and the fibril structures of the filament channel, resulting in the formation of bi-directional flows along the lower edge of the filament.
At about 06:22 UT, two hours after the C3.0 flare, brightenings started to appear at the opposite-polarity magnetic flux region once more (panels ({\it b}) and ({\it j})).
When contours of the \emph{RHESSI} HXR 6-12 (cyan) and 12-25 (blue) keV sources were superimposed on a 304 \AA\ image (panel ({\it i})),
it became clear that the \emph{RHESSI} HXR sources were spatially coincident with the EUV brightening.
Subsequently, it is clear from the NVST  $H_{\alpha}$ images (panels ({\it c})\sbond ({\it f})) that a cluster of dark fibril structures
was lifted from the brightening region and then merged with the filament channel. By scrutinizing the AIA observations (panels ({\it k})\sbond ({\it i})),
we found that bright material, which probably implies heated plasma, was also injected into the filament channel at the same time.
As a consequence, a significant new filament was gradually formed (panels ({\it g}) and ({\it n})).
The reformed filament underwent a series of dynamic evolution from about 07:00 UT to 09:00 UT
(panels ({\it f})\sbond ({\it h}) and (panels ({\it m})\sbond ({\it n})). Meanwhile, the filament material underwent
a to-and-fro movement along its axis. The material of the filament firstly moved to the eastern ends of the filament
(as indicated by the curved arrow in panel ({\it g})) and then moved to the western ends of the filament (as indicated by the curved arrow in panel ({\it h})),
and then back again. The movement of the filament plasma and the dynamic evolution of the filament
can be clearly seen on the associated animation. Finally, the filament gained an inverse S-shape,
with its western ends at the positive ends of the erupted filament and rooted in P, and with its eastern ends at
the negative ends of the emerging magnetic flux and rooted in n (panel ({\it p})). By considering the shape of the filament
and the magnetic field polarity around it, the chirality of the filament was identified as dextral \citep{mar98},
which is in line with the erupted one. These observations provide indisputable evidence that the reformation of the filament
is caused by magnetic reconnection between the emerging magnetic flux and the remaining filament channel
underneath the erupted filament.

\subsection{Photospheric Magnetic Field Evolution}
Since continuous EUV brightenings occurred at the opposite-polarity magnetic flux region,
the underlying photospheric magnetic field evolution in this region should contain the key clues to reveal the eruption and formation mechanisms of the filament.
Through checking the HMI data, it is found that significant flux emergence commenced in the northeast of the negative sunspot N.
This is shown by the HMI vertical images in Figure 6({\it a})\sbond ({\it d}). The flux emergence has been observed since the start of
the observation at 20:00 UT on 2013 July 14. It persisted for the following seven hours and ceased at about 03:00 UT on 2013 July 15.
As is common behavior for emerging magnetic flux, the positive and negative poles of the emerging magnetic flux moved away from each other
(panels ({\it a})\sbond ({\it b})). The positive pole, p, moved northwestward and then approached the negative sunspot n1 (panels ({\it a})\sbond ({\it b})),
which is firstly located at the northwest of N and gradually moved northeastward. As a result, p collided and cancelled with n1 and then the area of p
and n1 reduced simultaneously (panels ({\it b})\sbond ({\it d})). However, the negative pole moved southeastward and partly merged with N and partly moved
to the east of N, forming several negative penumbral sunspots n (panels ({\it a})\sbond ({\it d})).
The temporal evolution of the negative and positive magnetic flux in the cancellation area are shown
by the blue and red curves in Figure 7, respectively. One can see that the positive flux persistently increased from
20:00 UT on 2013 July 14 to about 01:00 UT on 2013 July 15, while the unsigned negative flux persistently increased from 20:00 UT on 2013 July 14
to about 00:00 UT on 2013 July 15. The increase in the positive flux is contributed to the emerging magnetic flux moving into the red calculation box.
However, the increase in the unsigned negative flux is due to the movement of the negative sunspot n1 into the blue box (as shown in Figure 6({\it d})).
It is visible on Figure 7 that the obvious decrease of the positive flux and the unsigned negative flux actually happened from about 01:00 UT on 2013 July 15
and continued until the end of the observation.
During this time interval, the recurrent brightenings occurred, the filament erupted and reformed.

It is notable that the transverse fields, which are represented by the short red and blue arrows with the arrow length proportional to the relative field strength
and the alignment parallel to the field direction, at the flux cancellation sites were changed and enhanced significantly ( panels ({\it a}) and ({\it c}) of Figure 6).
Before the flux cancellation, there were connection of transverse fields among p, N, and n ( panel ({\it a})), which distinctly confirms that the negative pole of the
emerging flux partly mixed with N and partly formed the negative penumbral sunspots n. However, at this moment, connection of transverse fields between
p and n1 was not identified ( panel ({\it a})). The lack of connection between p and n1 indicates that p and n1 belong to two topologically unconnected
magnetic systems. With the occurrence of the recurrent EUV brightenings associated with the flux cancellation, it is evident that sheared transverse fields,
which are rooted at p and connected to n1, appeared ( panel ({\it c})). This means that new connectivity between the cancelling flux p and n1 were established
during the magnetic reconnection between the newly emerged positive fluxes p and its nearby negative fluxes n1. As suggested by \citet{wang93} and \citet{zhang01}
that the appearance of sheared transverse fields in between the cancelling opposite polarities may imply that the opposite polarities in the cancelling magnetic features
are not the footpoints of a single flux loop, but the footpoints from two separated loops. These observations suggest that magnetic reconnection occurred in the
lower atmosphere, and the flux cancellation should be the result of magnetic reconnection in the lower atmosphere \citep{wang93,yang18}.

To further reveal the detailed evolution of the photosphere during the flux emergence and cancellation, the HMI continuum intensity (Figure 6({\it e}))
and the high resolution NVST TiO images (Figure 6({\it f}) \sbond ({\it h})) were also checked. At the flux emergence stage,
it is obvious that a series of penumbral fibrils, which apparently connect the positive spot p with the main negative sunspot N, appeared (Figure 6({\it e})\sbond ({\it f}).
As a result, the positive sunspot p and the main negative sunspot N share a common penumbra, indicating that a $\delta$ sunspot is formed.
In the course of the recurrent EUV brightenings and associated flux cancellation, these penumbral fibrils were sheared (Figure 6({\it g}).
In particular, the most striking features in the flux cancellation stage was that another significant new penumbra consisting of alternating dark and bright fibrils
gradually appeared in between the positive spot p and the negative spot n1 (Figure 6({\it g}) \sbond ({\it h})).
Comparing panel ({\it h}) with panel ({\it c}), it is clear that the sheared penumbral fibrils connect p to n1.
The emerged positive spot p and the negative spot n1 also share a common penumbra, meaning that another $\delta$ sunspot
is created during the flux cancellation. The rapid formation of a $\delta$ sunspot associated with flux cancellation
has been reported by \citet{wang13}. They interpreted their observations as being due to the eruption of a flux rope following
flux cancellation at the PIL, during which the re-closed magnetic arcades are pushed down toward the solar surface to form the new penumbra.
Generally, $\delta$ sunspots are believed to be formed by the emergence of a flux rope from the convection zone \citep{linton99,fang15,tak15,kni18}
or the interaction of two emerging flux ropes \citep{tor14,jou18}. In this observation, the first $\delta$ sunspot is directly formed
by the flux emergence. However, the second one is the result of the interaction between the emerging magnetic flux and its nearby magnetic fields.
The formation of the second $\delta$ sunspot accompanied with continuous brightenings and associated flux cancellation.
The gradually appearance of the penumbral fibrils in between p and n1 is direct and compelling evidence that magnetic connectivity
between p and n1 were gradually established by the magnetic reconnection in the lower atmosphere. The reconnection in the lower atmosphere
may result in the formation of new sheared magnetic arcades near the solar surface to form the penumbra. These observations also indicate that
the flux emergence and its driven flux cancellation may responsible for the development of the magnetic $\delta$ configuration of the AR.

\subsection{Magnetic Field Configuration}
With the aid of the photospheric vector magnetograms observed by the \emph{SDO}/HMI,
we carried out an NLFFF extrapolation to reconstruct the coronal magnetic field
of the AR at 01:00 UT (before the eruption of the filament) and 06:12 UT (just before the reformation of the filament) on 2013 July 15.
The results are presented in Figure 8. Different from many previous studies \citep{guo10,yan15,yang16b,zou16,hou18}, we detected no flux rope
or sheared magnetic field lines that are capable of supporting the concerned filament. However, the emerging magnetic fields (as indicated by the yellow lines)
and the overlying left-skewed coronal field lines (as indicated by the green lines) are well reconstructed by the NLFFF extrapolation.
It is clear that the emerged magnetic fields are sheared. In particular, a bundle of sheared field lines (as delineated by the white lines),
which connect P to n1, are existed before and after the filament eruption. From Figures 6(a) and Figure 8,
we can see that these sheared field lines
are found to match strikingly well with the $H_{\alpha}$ filament channel. Thus, we deduce that the magnetic structure
of the filament channel should be the sheared arcades and the filament channel is indeed lying below the filament.
The extrapolated results may reinforce our ideas that the first stage of interaction before the filament eruption occurred between
the emerged magnetic fields and the filament channel beneath the filament, and the reconnection between the emerged magnetic fields
and the filament channel resulted in the reformation of the filament. Moreover, the reconnection between the emerged magnetic fields
and the overlying left-skewed coronal loops can lead to the formation of a set of longer left-skewed coronal loops connecting p1
to N or n and a set of short loops connecting p1 to n1. The longer left-skewed coronal loops may manifest itself as the bright fan-like loops
and the hot channel, whereas the short loops may manifest itself as the bright sheared loops (see Figure 3({\it g}) \sbond ({\it h}))
in the corona and as sheared penumbral fibrils in the photosphere.

\section{Conclusion and Discussion}
Taking advantage of the high-resolution observations from NVST and \emph{SDO}, we study in detail the eruption and then reformation of a filament
caused by newly emerging magnetic flux, which emerged close to the negative ends of the filament in NOAA AR 11791.
Due to the flux emergence and its driven flux cancellation, the AR developed a magnetic $\delta$ structure.
Driven by the emerging magnetic flux, the newly emerged positive fluxes moved toward and cancelled with its nearby negative fluxes,
where the negative ends of a filament channel underneath the filament and a bundle of left-skewed coronal loops overlying the filament were rooted.
Supplemented by an NLFFF extrapolation, the sheared magnetic field lines of the filament channel and the emerged magnetic fields are extrapolated.
Unambiguous indications of magnetic reconnection between the canceling flux patches are identified, including the appearance of sheared transverse fields
in between the canceling opposite polarities, the formation of a $\delta$ sunspot, as well as the occurrence of EUV brightenings repeatedly
at the flux cancellation sites. By tracking the evolution of the brightenings, we find two distinct stages of reconnection prior to the rise of the filament.
The first stage of reconnection occurred between the emerging magnetic fields and the fibril structures of the
filament channel, resulting in the recurrent EUV brightenings and the bi-directional flows along the axis of the filament.
The emerging magnetic fields further interacted with the left-skewed coronal loops, indicating the second stage of reconnection,
which produced the recurrent EUV brightenings and associated bright fanlike structures striding the filament and stemming from the source region
of the EUV brightenings. In particular, during the final episode of the second stage of reconnection, a remarkable sigmoid structure was formed
and lifted up together with the filament. These observational features imply a tether-weakening reconnection, which diverted
the negative footpoints of the left-skewed coronal loops to the negative footpoints of the emerging magnetic flux and resulted in
lengthening the left-skewed coronal loops. Therefore, the magnetic tension force that restraining the filament was reduced, thus
the filament system was destabilized and began to rise. Subsequently, the filament and the sigmoid structure erupted together and produced a CME.
After the filament eruption, the emerging magnetic fields continued to reconnect with the filament channel. A bundle of dark fibril structures
was lifted from the flux cancellation region and then merged with the filament channel. At the same time, hot plasma was injected into the filament channel.
As a consequence, a new filament was formed. Our observation suggests that magnetic reconnection driven by
the emerging magnetic flux plays not only a vital role in the triggering mechanism of filament eruption, but also in the formation of filament.

Emerging magnetic flux has been served as a strong catalyst for the filament eruption onset. Previous observations and simulations have suggested
that reconnection-favored emerging magnetic flux are tightly corrected with filament eruption \citep{fey95,wang99,cs00,dac18}.
As magnetic reconnection happened between the emerged magnetic fields and the magnetic arcades overlying the filament,
it is called as tether-weakening reconnection by some authors \citep{moo92,nag07,ster07,ster14}.
\citet{cs00} carried out a two-dimensional magneto-hydrodynamic numerical simulations and found that when
reconnection-favored emerging magnetic flux emerges within the filament channel,
it cancels the magnetic fields below the filament, resulting in the rise of the filament due to loss of equilibrium,
and when reconnection-favored emerging magnetic flux appears on the outer edge of the filament channel,
it reconnects with the coronal loops overlying the filament, which can divert these loops sideways or to great height,
thus reduces the magnetic tension that restraining the filament, destabilizes the filament system, and then leads to its rise.
In our observations, the emerged magnetic flux is reconnection-favored,
as defined by \citet{fey95}, and it is close to the negative ends of the filament.
The emerged magnetic fields firstly reconnected with the filament channel underneath the filament.
However, during this stage of reconnection, our observations show that the filament is stable,
and the filament system is not destabilized by the reconnection. The rise of the filament happened after a series of reconnection that
occurred between the emerged magnetic fields and the left-skewed coronal loops overlying the filament. More precisely,
the rise of the filament was accompanied with the formation of a remarkable sigmoid structure during the final episode of the reconnection.
In the course of the expansion and eruption of this sigmoid structure, two dimming regions
gradually developed at its footpoints region. In particular, this sigmoid structure can only be seen on the AIA 94 and 131 \AA\ images. Based on these
observational results,  we speculate that this sigmoid structure should be the EUV hot channels reported by
\citet{zhang12}. The magnetic reconnection occurred between the emerged magnetic fields
and the left-skewed coronal loops produced two set of new magnetic field lines: a set of short loops that connects the emerged positive fluxes p
to the negative footpoints of the left-skewed coronal loops n1 and a set of new long left-skewed loops that connects the positive footpoints
of the left-skewed coronal loops p1 to the emerged negative fluxes N or even n. In our observation, the occurrence of the recurrent EUV brightenings
at the flux cancellation sites and the formation of the $\delta$ sunspot in the photosphere are indicative of the formation of the short loops.
The formation of the bright fanlike structures, which strides over the filament and stems from the source region of the EUV brightenings,
and the sigmoid structure are solid evidence that new long left-skewed loops were formed. In this study, the presented observations
are very similar to the simulation of \citet{cs00}. Based on these observational results,
we suggest that the rise of the filament is likely triggered by
the reconnection between the emerging magnetic flux and the magnetic arcades
overlying the filament. Once the filament starts to rise, the overlying arcades
should be stretched, a current sheet should be formed in the arcades below the rising filament,
reconnection then would set in the current sheet, and the following reconnection in the current sheet
could cause the filament to erupt to form a CME \citep{cs00,lf00}.

The formation of AR filaments are tightly related to emerging magnetic flux. Previous observations \citep{oka09,mact10} and
simulations\citep{arch08,fan09,arch14} have demonstrated that the emergence of a flux rope or sheared field lines combined
with magnetic reconnection, which driven by shearing and rotational motions during the flux emergence,
are key elements in the subsurface model of the filament formation.
When a flux rope emerged underneath a filament, the reconnection between them can create a new filament \citep{oka09,mact10,bu16}.
As emerged sheared field lines, which lie above the emerging tube axis, impacted by shearing motions and rotational motions, the reconnection of
them can also lead to the formation of a flux rope  \citep{arch08,fan09,arch14}. Our observation provides a unique perspective on
the formation of a filament caused by magnetic reconnection between the emerged magnetic field lines and a filament channel. In this study, we found that
the magnetic fields emerged close to the negative ends of a filament channel. After the filament erupted, the filament channel underneath the erupted
filament became visible. The filament channel consists of sheared fine chromospheric fibrils that
connecting P to n1 and can be effectively identified by the NVST high resolution $H_{\alpha}$ image (Figure 5({\it a})).
The emerged magnetic fields directly reconnected with the sheared fine chromospheric fibrils
of the filament channel, resulting in the EUV brightenings appeared at the flux cancellation sites and a cluster of dark fibril structures was lifted from
the flux cancellation region and merged with the filament channel. Furthermore, hot plasma was injected into the filament channel at the same time.
As a consequence, the filament reformed at the position of the erupted one.
These observational features are unambiguous evidence that magnetic reconnection  between the emerged magnetic fields
and the filament channel directly leads to the reformation of the filament. This observation indicates
that both emerging magnetic flux and its driven magnetic reconnection are indispensable elements
in the formation of AR filament.

Based on many observations and simulations, \citet{mac10} have concluded that there are there promising models,
including the injection model, the levitation model, and the
evaporation-condensation model, accounting for the formation of cool dense plasma in the filament.
In the injection model, through a series of magnetic reconnection processes, chromospheric plasma is injected into the filament channel
in the form of jets or upflows \citep{wang99,chae03,liu05,zou16,wang18,tian18}. In this model, magnetic reconnection is very likely
to occur when an emerging magnetic flux encounters an empty filament channel \citep{liu05,wang18}. The levitation model proposed that
the cool dense plasma of the filament can be directly lifted from the photosphere by emerging or reconnected field lines \citep{rus94}.
The evaporation-condensation model proposed that evaporated flows from the chromosphere can lead to radiative condensation
in the coronal loops to form the filament \citep{ant99,kar14,xia16}. However, in this model, in order to drive
the evaporated flows from the chromosphere, there needs a strong steady artificial heating at the footpoints of the coronal loops.
Recently, \citet{kan15,kan17,kan18} demonstrated that the change of the magnetic topology in a coronal magnetic field via magnetic
reconnection can trigger radiative condensation, which results in filament formation. More recently, \citet{li18} observed the coronal condensation
during the magnetic reconnection between a system of open and closed coronal loops. They pointed out that magnetic reconnection
and thermal evolution must be treated together and plasma condensation may naturally arise during the magnetic reconnection process.
In the present study, the formation of the filament may be the result of more then one mechanism at work.
Initially, photospheric plasma should be lifted to chromospheric heights by the emerged magnetic field lines,
and then heated and ejected into the filament channel via magnetic reconnection
between the emerging magnetic fields and the filament channel. As triggered by the reconnection,
the injected hot plasma should be condensed in the magnetic structure of the filament \citep{kan15,kan17,kan18}.
However, the real and detailed formation process of the cool dense plasma in the filament
must be more complicated than we thought and needs further investigation.

\acknowledgments
We thank the anonymous referee for constructive comments and suggestions that have improved the quality of the manuscript.
We also thank Jiayan Yang, Yi Bi, Junchao Hong, Haidong Li, Zhe Xu, and Hechao Chen for useful comments, as well as Rui Wang,
and Xiao Zhou for helpful discussions.
The data used in this paper are courtesy of the NVST, \emph{SDO}, \emph{RHESSI}, \emph {SOHO} and \emph{GOES} science teams.
This work is supported by the Natural Science Foundation of China, under grants 11703084, 11633008, 11873088, and 11573012;
the CAS ``Light of West China" Program; the Open Research Program of the Key Laboratory of Solar Activity of Chinese Academy of Sciences (KLSA201809);
the CAS grant ``QYZDJ-SSW-SLH012"; and the grant associated with the Project of the Group for Innovation of Yunnan Province.

\newpage
\begin{figure}
\epsscale{1.}
\plotone{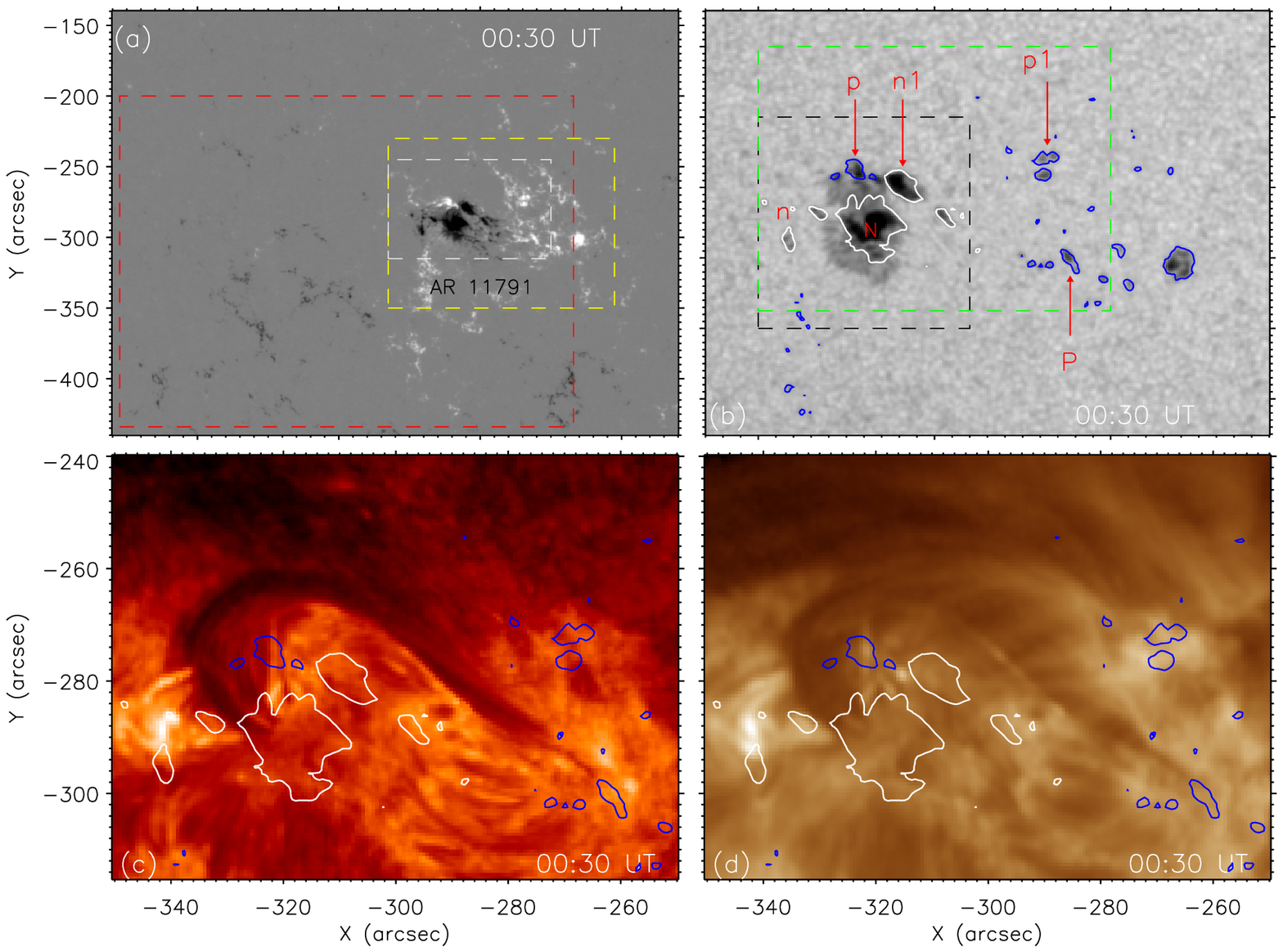}
\caption{Overview of the NOAA AR 11791 and the filament at 00:30 UT on 2013 July 15. Panel ({\it a}) shows the \emph{SDO}/HMI vertical image.
Panel ({\it b}) shows the HMI continuum intensity image. Panels ({\it c}) and ({\it d}) display the cotemporal AIA 304 \AA\ and 193 \AA\ images
showing the general appearance of the filament. Iso-Gauss contours of $\pm 600$ G are superposed by blue and white lines in panels ({\it b})\sbond({\it d}).
In panel ({\it b}), the letters ``n," ``n1," and ``N," register the negative magnetic flux patches,
while ``p," ``p1," and ``P" denote the positive magnetic flux patches.
The field of view (FOV) of panel ({\it b}) is outlined by the yellow dotted rectangle in panel ({\it a}).
The green dotted box marks the FOV of panels ({\it c}) and ({\it d}), the red dotted box denotes the FOV of Figure 3({\it i}),
the white dotted box shows the FOV of Figure 5, and the black dotted box indicates the FOV of Figure 6({\it a})\sbond({\it e}).
}
\end{figure}

\begin{figure}
\epsscale{1.}
\plotone{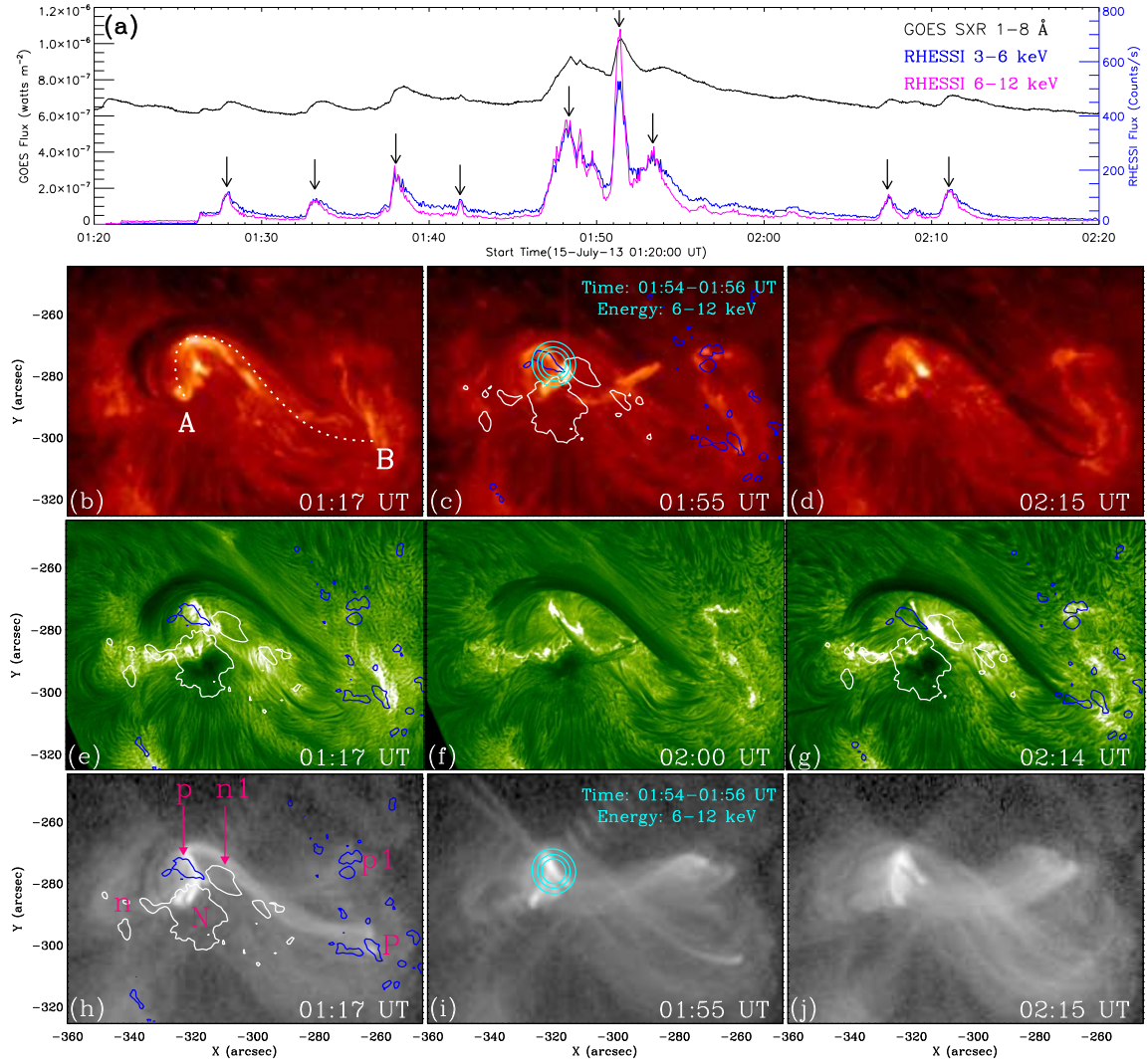}
\caption{Evolution of the filament and the recurrent EUV brightenings at the flux emergence sites.
Panel ({\it a}) displays the \emph{GOES} SXR flux at 1\sbond8 \AA\ (black) and added the
\emph{RHESSI} HXR light-curves at the energy ranges of 3\sbond6 keV (blue) and 6\sbond12 keV (pink) added{,respectively}.
({\it b})\sbond({\it d}) are AIA 304 \AA\ images, ({\it e})\sbond({\it g}) are NVST $H_{\alpha}$ line center images,
and ({\it h})\sbond({\it j}) are AIA 94 \AA\ images. Simultaneous HMI magnetograms are overplotted on panels ({\it c}), ({\it e}), ({\it g}),
and ({\it h}) as blue/white contours for positive/negative polarity, with contour levels of $\pm 500$ G.
The cyan contours (60\%, 80\%, and 90\% of the maximum X-ray flux) in panels ({\it c}) and ({\it i}) represent the \emph{RHESSI} 6-12 keV source.
The dotted line ``AB" in panel ({\it b}) marks the slit position of the time slice shown in Figure 4({\it a}).
An animation of panels ({\it b})\sbond({\it d}) and ({\it h})\sbond({\it j}) is available. The animation has 4 s cadence,
covering 00:30 UT to 02:30 UT.
(An animation of this figure is available.)}
\end{figure}

\begin{figure}
\epsscale{1.}
\plotone{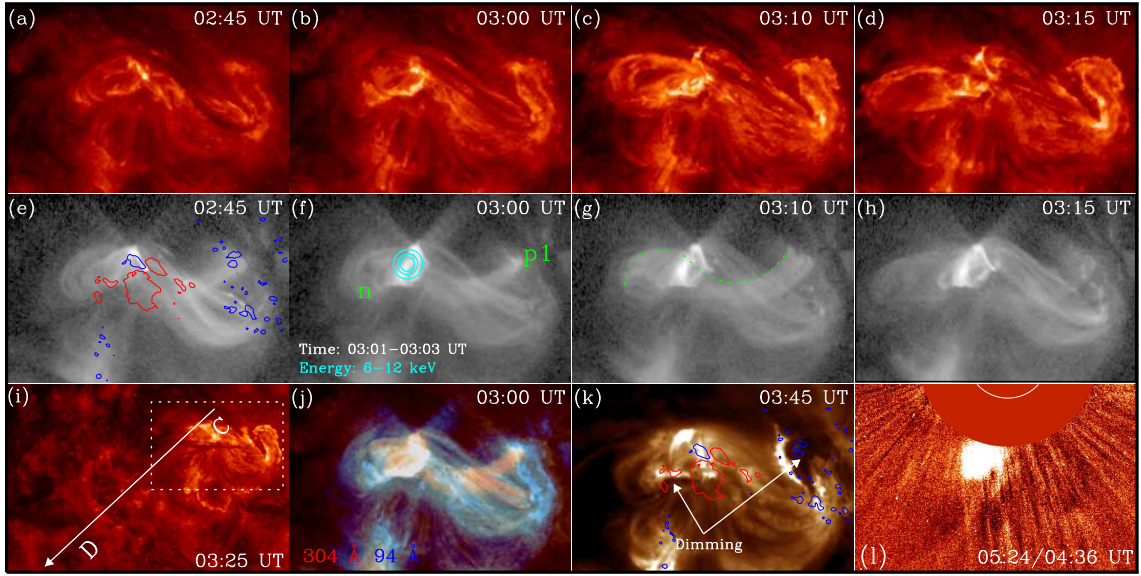}
\caption{Sequence of AIA 304 \AA\ (panels ({\it a})\sbond({\it d}) and ({\it i})) and 94 \AA\ (panels ({\it e})\sbond({\it h})) images
showing the eruption of the filament. ({\it j}) Composite image of the AIA 304 \AA\ and 94 \AA\ passbands displaying the overlying
sigmoid structure and the rising filament. AIA 193 \AA\ (panel ({\it k})) image showing the
associated coronal dimming (as indicated by the arrows). ({\it l}) LASCO/C2 white-light base-difference image exhibits the CME.
Likewise, the cyan contours (60\%, 80\%, and 90\% of the maximum X-ray flux) in panel ({\it f}) represent \emph{RHESSI} 6-12 keV source.
Iso-Gauss contours of $\pm 500$ G are superposed by blue and white lines in panels ({\it e}) and ({\it k}).
The arrow ``CD" in panel ({\it i}) marks the slit position of the time slice shown in Figure 4({\it b}),
while the green dotted line in panel ({\it g}) outlines the expanding sigmoid structure.
The dashed box in panel ({\it i}) shows the FOV of panels ({\it a})\sbond({\it d}), ({\it e})\sbond({\it h}), and ({\it j})\sbond({\it k}).
An animation of panels ({\it a})\sbond({\it i}) is available. The animation has 4 s cadence,
covering 02:30 UT to 04:30 UT.
(An animation of this figure is available.)}
\end{figure}

\begin{figure}
\epsscale{1.0}
\plotone{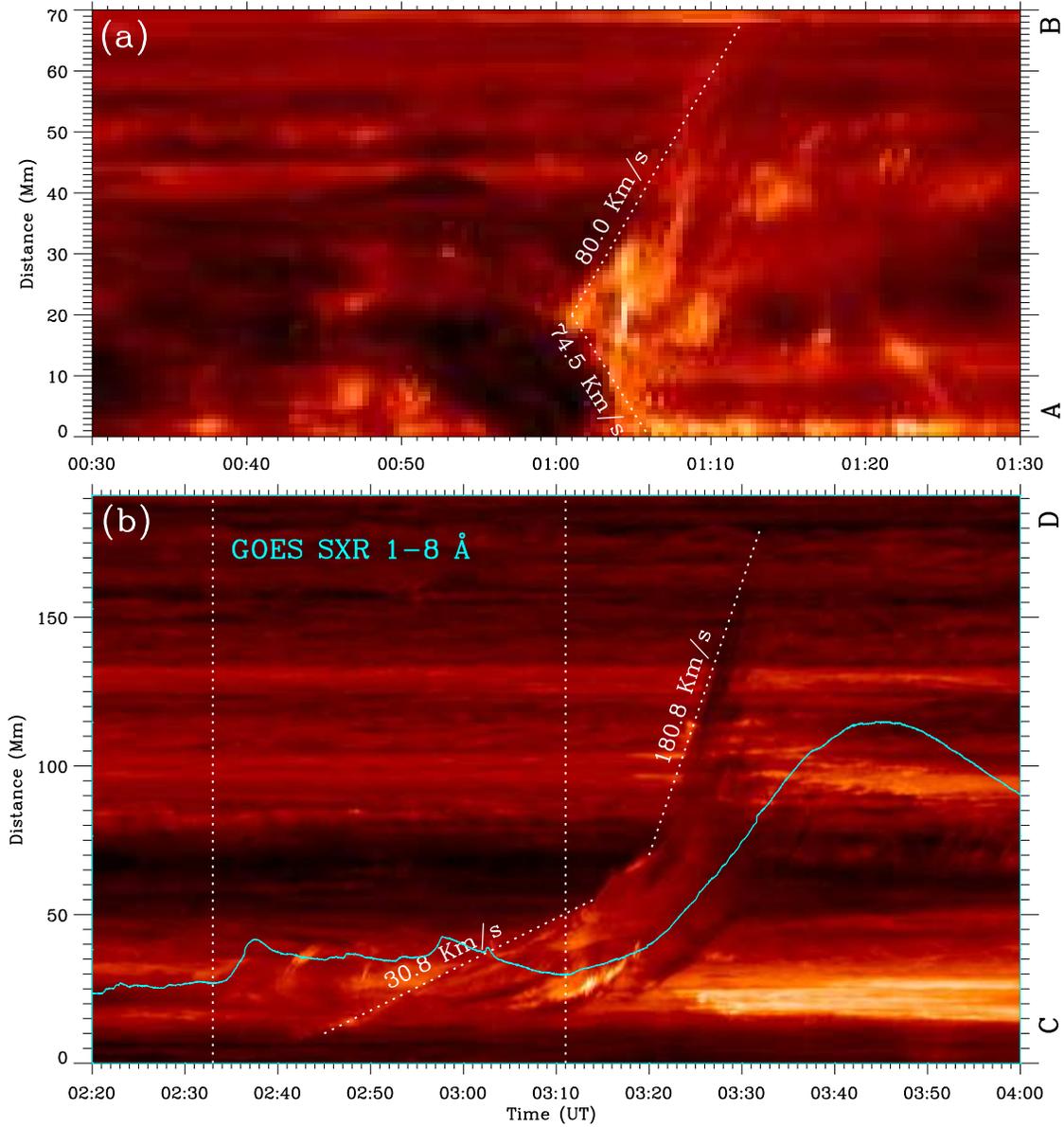}
\caption{({\it a})\sbond({\it b}) Time slices made from AIA 304 \AA\ images separately along the dashed lines AB and the arrow CD in Figure 2({\it b}) and Figure 3({\it i}).
The declining dashed lines in ({\it a}) mark the linear fitting to the outer edges of the brightening,
while the declining dashed lines in ({\it b}) mark the linear fitting to the outer edges of the erupting filament.
The cyan curve denotes the \emph{GOES} SXR flux at 1\sbond8 \AA\ . The vertical dashed lines indicating the start time of the C1.0 and C3.0 flares, respectively.}
\end{figure}

\begin{figure}
\epsscale{1.}
\plotone{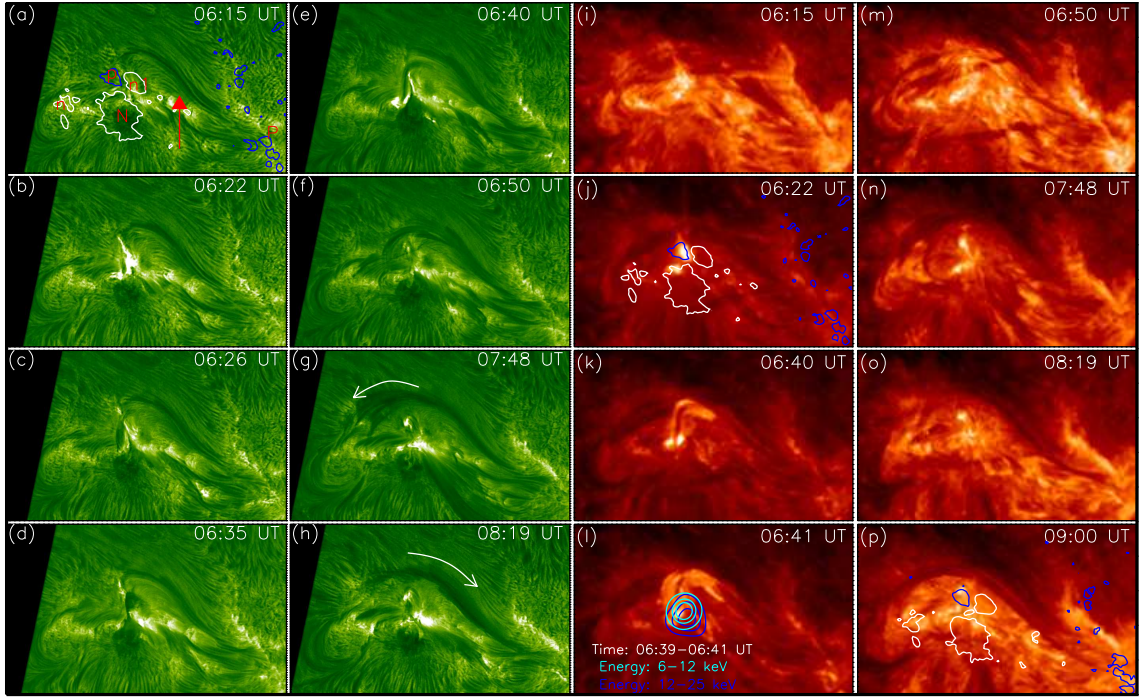}
\caption{Sequence of NVST $H_{\alpha}$ line center (panels ({\it a})\sbond({\it h})) and
AIA 304 \AA\ (panels ({\it i})\sbond({\it p})) images showing the reformation process of the filament.
Iso-Gauss contours of $\pm 500$ G are superposed by blue and white lines in panels ({\it a}), ({\it j}), and ({\it p}).
In panel ({\it a}), the red arrow points to a filament channel, which consists of fine fibril structures connecting P to n1.
The curved arrows in panels ({\it g}) and ({\it h}) denote the moving direction of the cool materials along the axis of the filament.
The cyan and blue contours (60\%, 80\%, and 90\% of the maximum X-ray flux) in panel ({\it i})
denote the \emph{RHESSI} 6-12 and 12-25 keV sources, respectively. The animation has 7 s cadence,
covering 06:10 UT to 08:20 UT.
(An animation of this figure is available.)}
\end{figure}

\begin{figure}
\epsscale{1.1}
\plotone{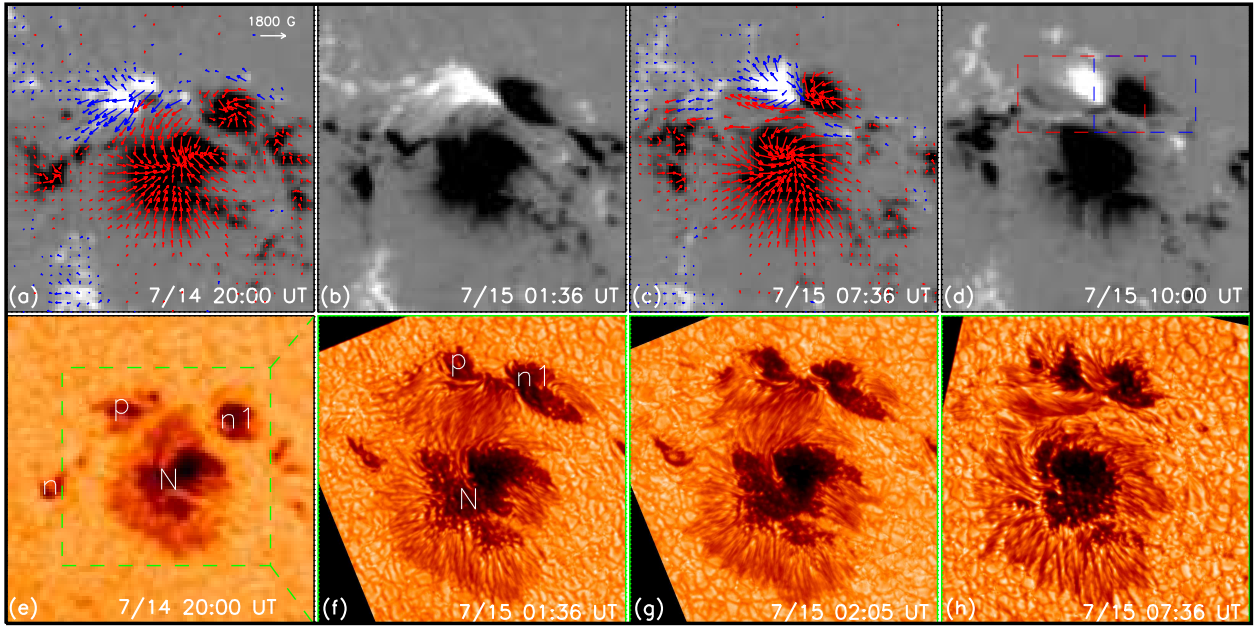}
\caption{\emph{SDO}/HMI vertical images showing the convergence and cancellation of opposite polarities at the region
covering the negative footpoint of the filament (panels ({\it a})\sbond({\it d})). HMI continuum intensity image (panel ({\it e}))
and NVST TiO images (panels ({\it f})\sbond({\it h})) present the evolution of the sunspots.
The green dotted box in panel ({\it e}) indicates the FOV of panels ({\it f})\sbond({\it h}).
The letters ``n," ``n1," ``N," and ``p" have the same meaning as in Figure 1({\it b}).
In panels ({\it a}) and ({\it c}), the red and blue arrows denote the horizontal magnetic field vectors,
which originate from a negative and positive longitudinal field where the magnitude exceeds 50 G.
The red and blue dotted boxes in panel ({\it d}) enclose the area are used to calculate the magnetic flux
from positive and negative polarities, respectively.}
\end{figure}

\begin{figure}
\epsscale{1.1}
\plotone{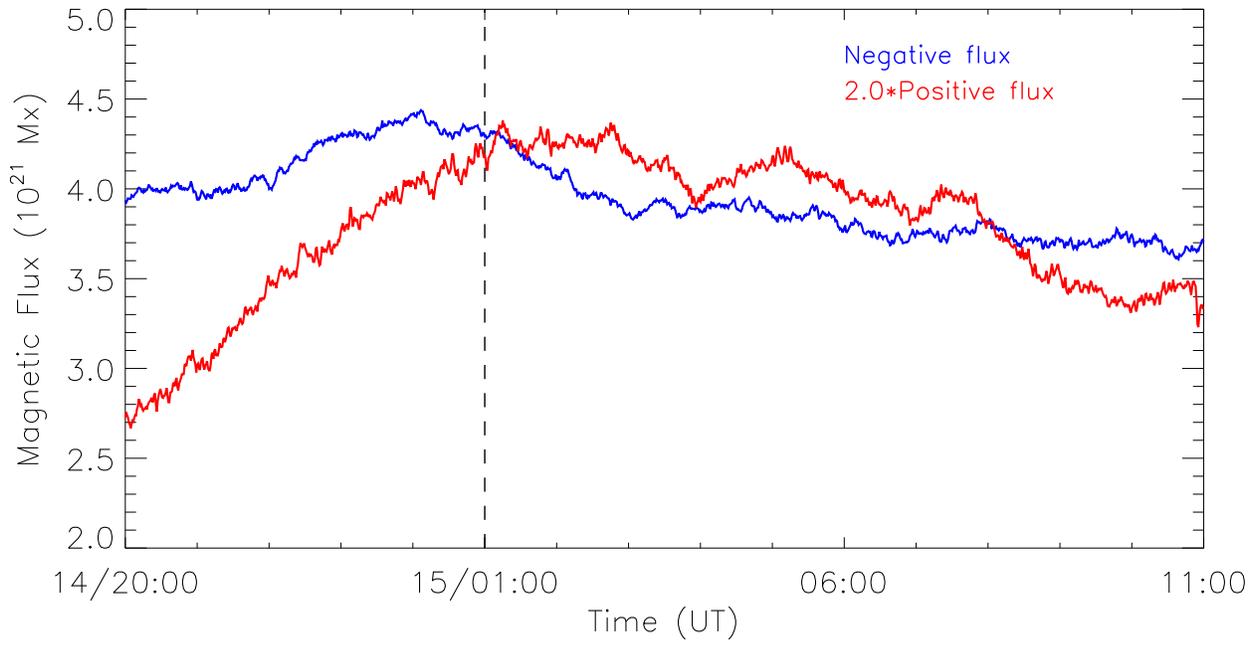}
\caption{Temporal evolution of the positive magnetic flux (red) and negative magnetic flux (blue)
in the red and blue dotted boxes shown in Figure 6({\it d}).
The vertical dashed lines denotes the time when the flux cancellation obviously occurs.}
\end{figure}

\begin{figure}
\epsscale{1.0}
\plotone{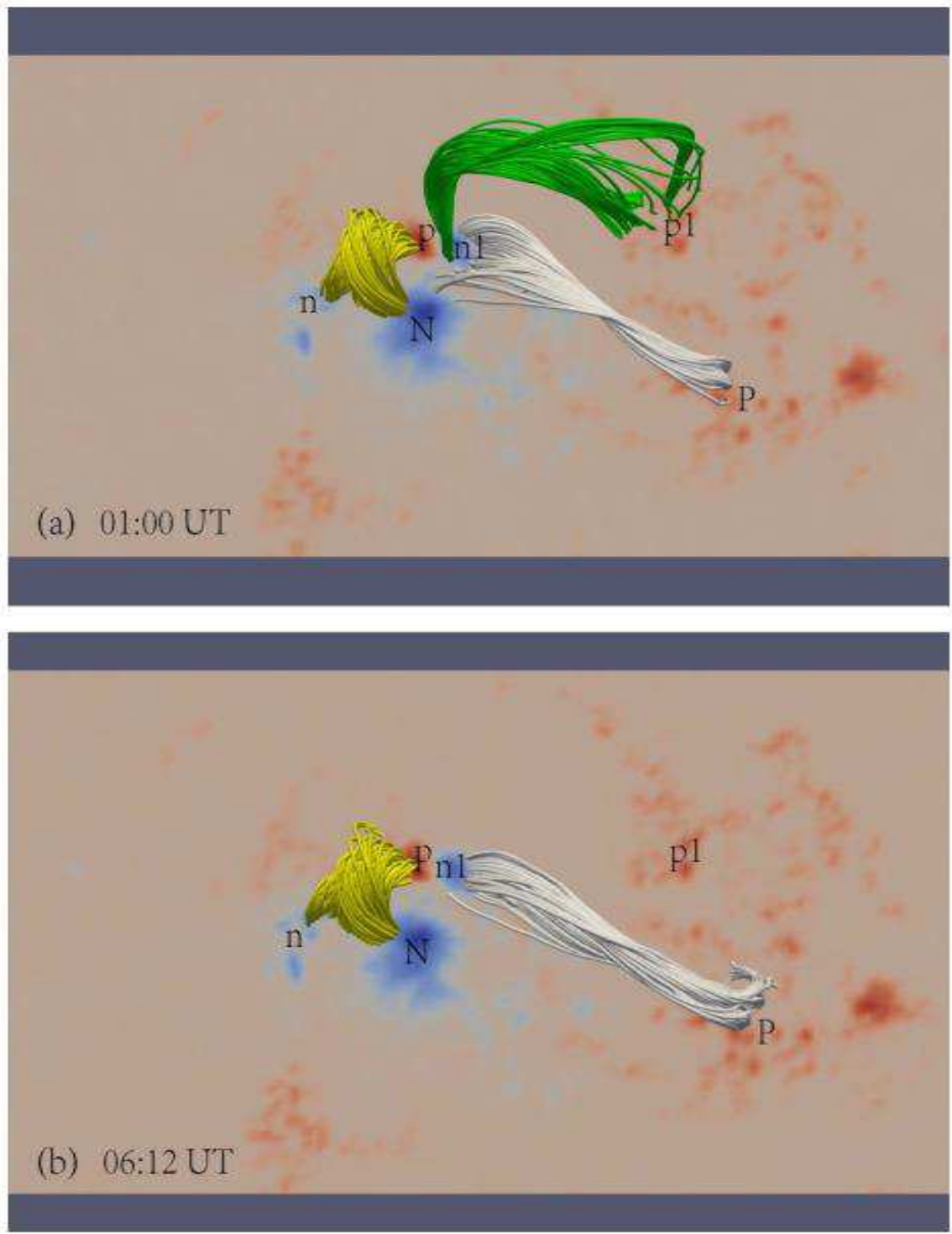}
\caption{HMI vertical field images ({\it a})\sbond({\it b}) superimposed with selected NLFFF lines.
The yellow lines in all panels represent the emerged magnetic field lines that rooted in P, N, and n.
The white lines in all panels indicate a bundle of sheared magnetic field lines  that rooted in P and n1.
Whereas the green lines in panel ({\it a}) are the overlying left-skewed coronal magnetic field lines that rooted in p1 and n1.}
\end{figure}


\end{document}